\documentstyle[12pt,a4,epsf]{article}
\begin{document}
\font\blackboard=msbm10 at 12pt
 \font\blackboards=msbm7
 \font\blackboardss=msbm5
 \newfam\black
 \textfont\black=\blackboard
 \scriptfont\black=\blackboards
 \scriptscriptfont\black=\blackboardss
 \def\bb#1{{\fam\black\relax#1}}
\newcommand{\z}{{\bb Z}}
\newcommand{\r}{{\bb R}}
\newcommand{\bc}{{\bb C}}
\newcommand{\be}{\begin{equation}}
\newcommand{\ee}{\end{equation}}
\newcommand{\ber}{\begin{eqnarray}}
\newcommand{\eer}{\end{eqnarray}}
\newcommand{\lp}{\left(}
\newcommand{\rp}{\right)}
\newcommand{\lk}{\left\{}
\newcommand{\rk}{\right\}}
\newcommand{\lc}{\left[}
\newcommand{\rc}{\right]}
\newcommand{\sT}{{\scriptscriptstyle T}}
\newcommand{\2}{\,\,2}
\def\a{\alpha}
\def\b{\beta}
\def\g{\gamma}
\newcommand{\se}{\section}
\newcommand{\tra}{\vec{p}_{\sT}}
\newcommand{\Z}{Z\left(\beta\right)}
\newcommand{\half}{\frac{1}{2}}
\thispagestyle{empty}

\begin{flushright}
\begin{tabular}{l}
FFUOV-98/02\\
{\tt hep-th/9803058}\\
\end{tabular}
\end{flushright}

\vspace*{2cm}

{\vbox{\centerline{{\Large{\bf FINITE TEMPERATURE MATRIX THEORY
}}}}}

\vskip30pt

\centerline{Marco Laucelli Meana, M.A.R. Osorio, and Jes\'{u}s Puente
Pe\~{n}alba
\footnote{E-mail address:
     laucelli, osorio, jesus@string1.ciencias.uniovi.es}}

\vskip6pt
\centerline{{\it Dpto. de F\'{\i}sica, Universidad de Oviedo}}
\centerline{{\it Avda. Calvo Sotelo 18}}
\centerline{{\it E-33007 Oviedo, Asturias, Spain}}

\vskip .5in

\begin{center}
{\bf Abstract}
\end{center}


We present the way the Lorentz invariant canonical partition function for
Matrix Theory as a light-cone formulation of M-theory can be computed. We
explicitly show how when the eleventh dimension is decompactified, the 
${\cal N} = 1$ eleven dimensional SUGRA partition function appears. 
We also provide a high temperature expansion which captures some
structure of the canonical partition function when interactions amongst
D-particles are on. The connection with the semi-classical computations
thermalizing the open superstrings attached to a D-particle is also
clarified through a Born-Oppenheimer approximation. Some ideas about 
how Matrix Theory would describe the complementary degrees of 
freedom of the massless content of eleven dimensional SUGRA are discussed.
Comments about possible connections to black hole physics are also made.
\newpage

\section{Introduction}

It has been intense the attention that the proposal known as Matrix
Theory \cite{bfss,intro} has attracted from the community of physicists working on the
field String Theory has become after the so called second string
revolution.  It seems that after a first boost of frenetic activity few
physical quantities have been computed so as to test to what extent we
can at least recover what we knew before from D-brane dynamics. At
present, it could be considered pretentious trying to get exact non
perturbative results from Matrix Theory. In Theoretical Physics, the
way out of this kind of situations is to wait for a more powerful
formulation of what we have at hand. Today's form of Matrix Theory has
some drawbacks. First of all the original formulation was one in the
IMF. This means that one should take a subtle
$N\longrightarrow\infty$ limit to recover Lorentz covariance. A
reformulation of the conjecture by using discrete light-cone
quantization (DLCQ from now on) seems to provide a meaning for the
computations at finite $N$. In DLCQ, it is the
light-like coordinate $x^-$ that gets compactified. In this form, the
conjecture would establish a correspondence between finite $N$ Matrix
Theory and the DLCQ of M-theory. Support for this correspondence has
been given in \cite{dlcq}.  Another related point is that the
present form of Matrix Theory is not a background independent one. A
list of Matrix Models for given backgrounds is in need and, up to now,
things do not go very fast to this respect. Something as modest as a
formulation of Matrix Theory for toroidal backgrounds is still lacking.
Consequently, Matrix Theory might be nothing more than another step
towards getting a more complete description of what in the past we
called String Theory. It is then not clear how far we will be able to
go.

It seems however that there are some ingredients that could show up as
advantages. After all we have a Hamiltonian and perhaps
this could be used to compute physical quantities that from their own
definition are adapted to that formulation. This would be the case of the canonical partition function,
$Z\left(\b\right)$. This work will be devoted to computing
this quantity, although the task will not be exactly accomplished. The
interest of the work will be that of getting a better knowledge of what
Matrix Theory Thermodynamics might be. From the very beginning it seems
clear that with $T=1/\b$ as the thermodynamical temperature we use to
define the canonical ensemble, the canonical partition function appears
to have only sense in the DLCQ picture because the IMF formulation of
the conjecture is actually throwing away energy configurations it seems
one should take into account at finite temperature.  We will focus on
the case in which all spatial dimensions are open with the exception of
$x^-$. One might argue that owing to the fact, also known in string
perturbation theory, that compact spaces increase the number of degrees
of freedom even to the extent of making difficult a formulation of a
Matrix Model for $T^{n}$ for $n\geq 6$, one should compute $\Z$ for a
$T^9$ space and get from here $\Z$ for $\r^9$. This prevention assumes
that this tremendous amount of degrees of freedom do not fully
disappear after decompactification. Such an attitude is based on the
extended misconception that a microcanonical description of
perturbative string theory gives a decompactification limit which is
different from that of the microcanonical picture gotten directly from
$\Z$ for an already open box (cf. \cite{us}). Then there is no a priori
reason to think that a formulation of Matrix Theory on a $T^9$ space,
if there is such a thing, would not give what we have now as its
decompactification limit. In fact, by computing the free energy of the
D-strings, by the way, we will check that the extra degrees of freedom
from the extra compact dimension decouple when its length goes to
infinity.

For the time being, in section 2, we will expose the technique to
compute $\Z$ for a theory which is formulated in the light-cone frame.
The main point is that if $T=1/\b$ can be treated as the temperature one
could measure with a thermometer, one has to admit that it has to
transform as the energy, i.e. $p^0$. In other words $\Z =
\mbox{Tr}\,e^{-\b p^0}$ is a Lorentz invariant quantity we are going to
compute knowing only a description in a light-cone frame. We will
present this computation in the DLCQ form of the Matrix Theory
conjecture. In section 3 we will study the limiting case in which one
has free (classical geometry) configurations and what happens when
off-diagonal terms (open stretching strings) are taken into account. In
section 4, we will study a classical system of matrices or what more
properly could be defined as the classical statistical mechanics
associated with Matrix Theory and we will relate it to Matrix Theory at
finite temperature by computing its quantum corrections. In section 5
we will perturbatively compute the partition function by a
finite temperature extension of the computation in
\cite{bb}. We will finally get out the conclusions in section 5.

\section{The canonical free energy from a light-cone frame description}

The canonical partition function for the simplest relativistic
system of a gas of particles of mass $m$ is by definition (see
\cite{enrial})

\be
\Z = \mbox{Tr}\, e^{-\b p^0}.
\label{barz}
\ee
This is an invariant quantity whenever we admit that the temperature
$T$ transforms as the energy, i.e. the zero component of the momentum.
This is in physical terms a consequence of the interpretation of the
temperature as measuring the average kinetic energy of the system.
If the system is known in the light-cone frame, one can use
the relation $p^0 = \frac{1}{2}(p^{+} + p^-)$ to write

\ber
\Z & = & \mbox{Tr}\, e^{-\half\b\lp p^+ + p^-\rp} = \nonumber\\
   &   & \mbox{Tr}_{p^+,\,\tra}\lc\,
   e^{-\half\b\lp p^+ + \frac{m^2}{p^+}\rp}\, \, 
           e^{-\b\frac{{\vec{p}_{\sT}^{\2}}}{2p^+}}\rc .
\label{lc}             
\eer

Let us perform the trace by first taking a very illustrative path. One
can compute a single particle partition function by performing the
trace in (\ref{lc}) using a basis $|p^+>\otimes\,\,|\tra >$ of
single particle momentum eigenstates. If the space volume goes to
infinity, we have

\be
\frac{L}{2\pi}\,\int_{0}^{+\infty}\,dp^+\,e^{-\frac{\b p^+}{2}}\,
e^{-\frac{m^2\b}{2p^+}}\,\frac{L^{d_{\sT}}}{\lp2\pi\rp^{d_{\sT}}}\,
\int\,d^{d_{\sT}}p\,\,e^{-\frac{\b\tra^2}{2p^+}}.
\ee
The change of variables $s=\frac{\b}{p^+}$ and an elementary integration
take us to

\be
\frac{L^{d-1}\b}{\lp2\pi\rp^{\frac{d}{2}}}\,\int_0^{+\infty}ds\,
s^{-1-\frac{d}{2}}\,e^{-\frac{\b^2}{2s}}\,
e^{-\frac{m^2 s}{2}}.
\label{m-b}
\ee
Where $d$ is the space-time dimension, $d=d_\sT + 2$, and
$L\longrightarrow +\infty$. The variable $s=\frac{\b}{p^+}$ is the
Schwinger proper time, and we have the well known proper time
representation of the Helmholtz free energy. Eq. (\ref{m-b}) can be
recognized as $-\b F_1\lp\b\rp\equiv Z_1\lp\b\rp$ for a particle of mass
$m$ and also as the same magnitude per degree of freedom for a quantum
field with particle excitations of mass $m$. In both cases one has
Maxwell-Boltzmann statistics.

To get $\Z$, i.e. the canonical partition function, a further
exponentiation and another sum are needed (see \cite{emar87}). The
relation, for particles obeying Bose statistics, is

\be
\ln{\Z} = \sum_{r=1}^{+\infty}\frac{Z_1\lp r\b\rp}{r}.
\label{canonical}
\ee
The exponentiation gives the multi-particle system and the sum over $r$
gives bosonic quantum statistics versus Maxwell-Boltzmann which
corresponds in fact to the $r=1$ term in the sum. So one has two
aspects which, in principle, are different. One can have a
multi-particle description but with Maxwell-Boltzmann statistics
instead of Bose or Fermi quantum statistics. This subtlety shows up
clearly as soon as one deals with the computation of the canonical
partition function for string perturbation theory in the light-cone
gauge. It is worth noticing here that the single-particle partition
function depends linearly on the open (infinite) volume and then the
multi-particle partition function depends exponentially on the volume.

Eq. (\ref{canonical}) is the result of computing $\Z$ as the trace of
$e^{-\b p^0}$ using a basis, in the Fock space, of multi-particle
eigenstates of the energy $p^0$ assuming Bose statistics. The rigorous
way of getting (\ref{canonical}) is assuming that our spectrum is
actually discrete. Physically this can be achieved by enclosing the
system in a finite volume box. One can then write

\be
\Z = \prod_{\vec{k}}\,\lp 1 \pm e^{-\b\omega_{\vec{k}}}\rp^{\mp 1}.
\ee
Here $\omega_{\vec{k}}$ is the energy, that is usually understood as a
dispersion relation and the signs distinguish bosons from fermions.
Taking the logarithm of both sides, one gets the Helmholtz free energy
as an infinite sum. It is from here that one gets the sum over $r$ by
expanding $\ln{\lp1\pm x\rp}$. Physically it is not necessary to
assume that the compactified theory is known if one makes the
hypothesis that, after taking the infinite volume limit that convert
the sum over discrete momenta into an integral, the remnant degrees of
freedom are those of the formulation of the system for an infinite box
which is supposed to be what is known.

In quantum field theory one can check that $\ln{\Z}$ can also be
obtained by a Euclidean path integral computation of the covariantly
defined theory on ${\bb R}^{d-1}\times S^1_\b$. After clarifying some
questions related to modular invariance, the same correspondence will
work for perturbative string theory, but a light-cone gauge computation
based on the transverse Hamiltonian does not produce a multi-string
partition function on its own. Let us explain what we mean. The
starting point is again Eq. (\ref{barz}) which can be rewritten as

\be
\Z = \frac{L}{2\pi}\,\int_0^{+\infty}\,dp^+\, e^{-\frac{\b p^+}{2}}\,
z_T\lp\frac{\b}{2p^+\a^{'}}\rp
\ee
Here, $z_T$ is defined as a transverse partition function given by

\be
z_T\lp\lambda\rp = \mbox{Tr}_{H_T}\, e^{-\lambda H_T}
\ee
with $H_T$ the light-cone gauge Hamiltonian associated with the
light-cone gauge (transverse) action that for the bosonic part reads

\be
S^{l.c.}_{\mbox{\tiny bosonic}} =
-\frac{1}{4\pi\a^{'}}\,\int\,d\sigma\,d\tau\, \partial_\a X^i\partial^\a
 X^i\,\,\,\,i=1\mbox{,...,}8 
\ee
$\lambda$ is a sort (it is a dimensionless quantity) of inverse
transverse temperature. With this action (and is fermionic part), $z_T$
can be computed by performing a Euclidean path integral with periodic
(anti-periodic) boundary conditions in the transverse temperature
$\lambda$ for the coordinates (32-spinor $S$) as fields in two
dimensions. The result one finally obtains for a string one-loop
computation is again a Maxwell-Boltzmann Helmholtz free energy
proportional to the infinite volume in nine dimensions; i.e. a single
string partition function $Z_1^{\mbox{\tiny string}}\lp\b\rp$ \cite{emar87}.

The supersymmetrical version of Eq.(\ref{canonical}) gives the free
energy for the multiple-string  gas. In short, it can be
written

\be
\mbox{ln}{\Z} = \sum_{r = \mbox{\tiny odd}>0}\frac{Z_1^{\mbox{\tiny
string}}\lp\b r\rp}{r} .
\label{susy}
\ee
Where one assumes that the single partition function includes the sum
over Maxwell-Boltzmann contributions coming from the bosonic and
fermionic degrees of freedom. It is worth noticing again that the
single-string partition function depends linearly on the infinite
volume while the multiple-string partition function exponentially
increases its free energy with the volume. In terms of the canonical
entropy, it is clear that the Helmholtz free energy has to depend
linearly on the volume as the entropy does.

Let us finally tackle our goal which is to get the form of the
partition function for the Matrix Model. One first notes that in the
DLCQ, $p^+$ is discrete because $x^-$ is a circle of radius $R$. From
an eleven dimensional point of view, what one would first write down
for the partition function of the model would be

\be
Z_1^{\mbox{\tiny matrix}}\stackrel{?} = \sum_{N=1}^{+\infty}\,
e^{-\b\frac{N}{2R}}\,\mbox{Tr}\, e^{-\b R H_{\lp N\rp}}
\label{tentative} 
\ee
where $H= R H_{\lp N\rp}$ is the Hamiltonian for the SYM with $U(N)$
symmetry given, with the gauge $A_0=0$, by

\be
H = R\,\mbox{Tr}\,\lp\frac{1}{2}\Pi_i^2
- \frac{1}{16\pi^2{\a^{'}}^3}\lc Y^i,Y^j\rc^2
-\frac{i}{2\pi{{\a^{'}}^{3/2}}}\,\pi\g^j\lc Y_j,\theta\rc\rp
\ee
Physically Eq. (\ref{tentative}) should be interpreted as the classical
partition function for a single eleven dimensional object. The question
is that we have to determine what is meant by the trace after the sum.
After knowing the features Matrix Theory have revealed since the
conjecture was formulated, it appears reasonable to us to study
particular configurations in order to try to identify the single object
one could use to make statistical mechanics the way it has been
presented here for particles, fields and perturbative strings. At first
sight one knows that in terms of ten dimensional physics one has
systems of $N$ free D0-branes and also bound states at threshold formed
by $N$ D0-branes. Let us then study the simplest picture which is that
of the  free (classical geometry) configurations in the Matrix
Model.

\section{Matrix Theory free configurations}

A simple analysis of the Hamiltonian we have written down shows that in
Matrix Theory there is no well defined meaning for the concept of
position interpreted as a way of defining a standard geometry for the
configuration space of the blocks we play with which are D0-branes. The
reason is well known and results from the natural thickness of
Dirichlet branes given by the fact that they have open strings on them
and actually at ultrashort distances stretching strings can glue
together two D0-branes, for example. These stretching strings actually
introduce some kind of fuzziness that makes a classical geometrical
interpretation for the configuration space of the branes an approximate
concept. If the $N\times N$ matrices we use to write $H_{\lp N\rp}$ are
diagonal, then a great simplification results because commutators
get null and the Hamiltonian reduces to the simple  form

\be
H_{\lp N\rp}^{\mbox{\tiny
free}}=\mbox{Tr}\,\lp\frac{1}{2}\Pi_i^2\rp ,
\ee
with each $\Pi_i$ being a diagonal $N\times N$ matrix. It is clear
because of the form of the Hamiltonian that the classical (geometry)
configurations correspond to a free theory. The states we want to perform the
trace over in Eq.(\ref{tentative}) are
$\lc\lp\otimes_{i=1}^{9}|p^i>\rp\otimes\,
|\tau>\rc_N$ with the subscript $N$ indicating that the states
are actually arranged in an $N$ component vector for a given $N$ and
$|\tau>$ carries the Pauli spin information.
With them, one can make the following guess

\be
 Z_1^{\mbox{\tiny free}}
\stackrel{?} = \sum_{N=1}^{+\infty}\,
e^{-\b\frac{N}{2R}}\,
\mbox{Tr}\, e^{-\b R H_{\lp N\rp}^{\mbox{\tiny free}}}
 = \sum_{N=1}^{+\infty}\,256^N
\frac{1}{N!}\,\frac{V_9^N}{{\lp 2\pi\rp^{9N}}}
\lp\frac{2\pi}{ R\b}\rp^{\frac{9}{2}N}e^{-\b\frac{N}{2R}} .
\label{tentativec}
\ee
Here the $N!$ factor comes from the permutation group which is the
discrete symmetry that survives from $U(N)$ after truncating the
matrices to the diagonal ones. $V_9$ is the spatial nine-dimensional
volume.

Now the next step would be that of substituting this single partition
function into Eq. (\ref{susy}) and getting $\Z$ after an
exponentiation. However there is something remarkable in
Eq.(\ref{tentativec}). It has been constructed as a single particle
function an as such should be identified with $-\b F_1\lp\b\rp$ and then
one would expect it to depend linearly on the volume and this is
not the case. In fact one can write

\be
Z_1^{\mbox{\tiny free}}
 \stackrel{?} = \mbox{exp}\lc{256
 \,\frac{V_9}{\lp 2\pi\rp^9}
 \lp\frac{2\pi}{ R\b}\rp^{\frac{9}{2}}e^{-\b\frac{1}{2R}}}\rc - 1\,,
 \label{second1}
 \ee
i.e., this partition function depends exponentially on the
nine-dimensional volume. This is simply the result of the fact that the
Matrix Model contains second quantization in itself in the sense that the
first quantization of an object in eleven dimensions gives the
description of ten dimensional multi-objects which are sets of free
D0-branes. Looking at it this way, Eq. (\ref{second1}) would
define the single object as one with single partition function

\be  
256\,\frac{V_9}{\lp 2\pi\rp^9}                    
\lp\frac{2\pi}{ R\b}\rp^{\frac{9}{2}}e^{-\b\frac{1}{2R}} .
\ee
The second quantization of this single object would produce composite
systems of free D0-branes. In other words, a given total momentum would
result from the sum of the individual momenta of each free D0-brane.
Then the special bound states at threshold, special because they are
bound states of $n$ D0-branes but with zero relative energy, are
missing and we are not taking into account that the given total
momentum can also be shared by several bound states at threshold. The
consequence is that we have to modify our single object to include
these states and that is easy. Our single object will be a sum over
partons of any positive RR charge. This is similar to the image of the single
fundamental string in the analog model as a collection of an infinite
number of fields with masses running from zero to infinity. At last,
the single object partition function will be

\be
 Z_1^{\mbox{\tiny free}}\lp\b\rp =
\sum_{k=1}^{+\infty}\,256\,\frac{V_9}{\lp 2\pi\rp^9}                            
 \lp\frac{2k\pi}{ R\b}\rp^{\frac{9}{2}}e^{-\b\frac{k}{2R}}
\label{object}
  \ee  
With this, we can now apply the recipes in Eq. (\ref{susy}) to get

\be
 Z^{\mbox{\tiny free}}\lp\b\rp
= \mbox{exp}\lc\sum_{r = \mbox{\tiny odd}>0}\frac{1}{r}
  Z_1^{\mbox{\tiny free}}\lp\b r\rp\rc .
\label{mul}
\ee
The relationship between $Z^{\mbox{\tiny free}}\lp\b\rp$
and our starting point in Eq. (\ref{tentativec}) is very clear. Eq.
(\ref{tentativec}) coincides with the $r=k=1$ term in 
$Z^{\mbox{\tiny free}}\lp\b\rp$ after expanding the
exponential. In other words it  is the multi-object
Maxwell-Boltzmann version of our system without counting the D0-brane
bound states at threshold. 

We see in (\ref{object}) that our single object is a superposition of
bound-states composed by $k$ D-particles. The dynamical degrees of freedom
of each parton does not contain the dynamics of its components, i.e. we do
not have $V_9^k$ or any permutation property of the $k$ D-particles as in a
standard gas with the same number of components; it only contains the
dynamics of the center of mass. This feature is interesting because it does
not prevent the interpretation of a given $k$-parton as a Schwarszchild
black hole (see \cite{blackholes1}). We will explain it in more detail in
section 5.

It is necessary to check the nature of our single object by identifying
into what  the single partition function turns after taking the
$R\longrightarrow +\infty$ limit that would open the eleventh dimension
up. It is an easy task to perform such limit because amounts to
converting the infinite sum over positive integers $k$ into an
integral. One has

\ber
 & &\sum_{k=1}^{+\infty} \,256\,\frac{V_9}{\lp 2\pi\rp^9}
  \lp\frac{2k\pi}{ R\b}\rp^{\frac{9}{2}}e^{-\b\frac{k}{2R}}
\,\,\,\stackrel{R\rightarrow +\infty}\longrightarrow  \nonumber\\
& &\frac{\lp 2\pi R\rp}{2\pi}
\int_{0}^{+\infty}\,dk\,256\,\frac{V_9}{\lp 2\pi\rp^9}
\lp\frac{2k\pi}{\b}\rp^{\frac{9}{2}}e^{-\b\frac{k}{2}} =\nonumber\\
& & 256\,\pi^{-11/2}\,\Gamma\lp\frac{11}{2}\rp\,V_{10}\,\b^{-10}\, ,
\eer
with $V_{10}$ the space volume in eleven dimensions. This, through Eq.
(\ref{mul}), gives $-\b F\lp\b\rp$ for the massless field content of ${\cal
N} = 1$ SUGRA in eleven dimensions. Our single object becomes an eleven
dimensional supergraviton when opening up the light-like dimension as, may
be, one could have expected from the very beginning. At finite temperature
we have to sum over the complete tower of longitudinal momentum modes, that
is, we have to take into account the contribution of all the finite-N Matrix
Models, this allows us to avoid the problem of relating the finite-N model
to the eleven dimensional supergravity \cite{kabtaylor}.

The question now is whether we can use all this analysis to treat
interactions among the D0-branes. At first sight one would say that all the
treatment for particles, fields, strings and the Matrix Model itself is only
accurate for free single objects. We know this is not true for fields and
strings. The proper time formalism is also accurate for including
interactions perturbatively through corrections $\delta m^2$ to $m^2$. One
only has to change the mass squared by the mass squared plus the loop
corrections. This can be done for strings in the analog model in which the
free energy is the sum over the field content of the string (cf.
\cite{moore}). We will actually see in section 5, that this can also be done
for the Matrix Model in the approach in which a perturbative expansion
around a fixed background is done. In some sense, what the light-cone
description of M-theory should provide for the massless content of ${\cal N}
= 1$ SUGRA is analogous to the massive tower of ten dimensional Planck
masses that promote the ten dimensional supergravities to the vibrational
modes of the full fundamental superstring theories. The Matrix Model
interactions through stretching strings are the sources for these massive
companions, although we do not see any a priori reason to believe that the
relevant degrees of freedom could be described as some kind of massive
quantum fields beyond this fixed background perturbation expansion approach.

%
%
%

\section{Statistical mechanics of $U(N)$ matrices}

We have already stressed as one  of the advantages of the present
formulation of Matrix Theory that it is based upon a relatively simple
Hamiltonian. This provides us with a different way to do our
computations without any need to do a hard path integral nor to
restrict ourselves to a particular perturbation series that strongly
depends on the background -the vacuum- we choose to expand around. At
least, as we shall see, it is possible to find global, non perturbative
features of the partition function that the field perturbation theory
cannot capture.

The Hamiltonian formulation fundamentally differs from the Lagrangian one 
in the mathematical objects that contain the complexity of the system. 
In the Lagrangian formulation, the function that defines the system is
the action, calculated over the trajectories or paths. This necessarily 
leads to a path integral if we want to calculate any physical quantity.
On the other hand, in the Hamiltonian formulation, the system is defined 
by a hermitian 
operator acting over wave functions. Physical magnitudes are calculated
through summations or integrals over real or Grassmann variables. 
The complexity is thus transported into the search for appropriate wave 
functions as well as their matrix elements with the operators that 
appear in the Hamiltonian.

Let us rewrite the transverse Matrix Theory Hamiltonian with every
$\hbar$ and $\a^{\prime}$ in it

\begin{equation}
H=R\,\mbox{Tr} \left(\frac{1}{2 \hbar}\Pi_i^2-\frac{1}{16 \alpha'^3 \hbar^2}
\left[Y^i,Y^j\right]^2
-\frac{i}{ 2\pi( \alpha' \hbar)^{3/2}} \pi \gamma^j
\left[Y_j,\theta \right]\right),              
\end{equation}
with the constraint
\begin{equation}
\pi=-i\sqrt{\frac{\hbar}{\alpha'}}\theta^t
\end{equation}
Again, we have chosen the gauge $A_0=0$. Besides, one has to impose another
condition that arises from the Gauss law. If we take the Arnowitt-Fickler
gauge in the $9+1$ theory, the condition after dimensional reduction to
$0+1$ is $Y^9_a=0$. The only consequence is that now, the bosonic index $i$
takes values from 1 to 8. This gauge election does not need ghost fields.
This is general for all the 'gluons' except for that related to the identity.
This one decouples from all the rest and does not even appear in the Gauss
constraints because trivially commutes with any other matrix
\footnote{In the special case of a  $0+1$ dimensionally reduced theory 
in the hamiltonian formalism, the Gauss constraints can be completely written 
in terms of the structure constants of the gauge group, that is
\cite{danfer}
$G_{\rho}=F^{\mu\nu}_{\rho} Y^{\mu}_j
\Pi^{\nu}_j-iF^{\mu\nu}_{\rho}\theta^{\mu}\theta^{\nu}.
$
Let us note that in the free case in which the gauge symmetry is broken to
$U(1)^N$ these conditions do not impose any extra constraint to the election
$A_0=0$ and so we keep nine physical scalar fields for each D-particle. This
justifies the calculation in section 3.}

More comments on the particular form of this Hamiltonian are also in need.
Four different operators appear: the fermionic and bosonic positions
$\theta$ and $Y$ and their conjugate momenta $\Pi$ and $\pi$. The
commutation relations among them are various. Firstly, all of them are
$N\times N$ matrices and therefore they obey the $U(N)$ algebra. Secondly,
they are either bosonic or fermionic creation operators of one-dimensional
world-line fields or, equivalently, operators with either real or Grassmann
numbers as eigenvalues from the ten dimensional point of view. So they
correspondingly commute or anti-commute according to their nature. Finally,
each coordinate and its conjugate momentum have canonical Heisenberg
commutation rules too.

Other noticeable characteristic of the Hamiltonian is the ubiquitous
appearance of $\hbar$ in every term. The physical origin of this
fact is that, from the eleven-dimensional point of view, the BPS
masses of the D0-branes are Kaluza-Klein energies and, therefore,
purely quantum mechanical.  Besides, let us remember that the Planck
length in any dimension has also got a purely quantum mechanical
nature. As it stands, it is impossible to say what is the limit of
the Hamiltonian as $\hbar \rightarrow0$, or even if there is any.

The first step we shall take is to expand the matrix
operators in terms of the generators of the $U(N)$ algebra in order
to avoid complications related to their matrix nature. This yields

\begin{equation}
H=\frac{R}{2\hbar}\Pi_\lambda^k\Pi_\lambda^k-\frac{R}{16\pi\alpha'^3}
F^{\mu\nu}_\rho F^{\tau\epsilon}_\rho Y^i_\mu Y^i_\tau Y^j_\nu Y^j_\epsilon
-\frac{i R}{{2\pi}\left(\hbar \alpha'\right)^{3/2}}
\pi^\mu Y_\sigma^i F^{\sigma \nu}_\mu \gamma_i \theta_\nu
\end{equation}
where $F^{\sigma\nu}_\mu$ are the structure constants of the gauge group. To
make the trace over the exponential of this Hamiltonian we have to choose a
base of wave functions. The most appropriate is formed with eigenvectors of
the $16N^2$ fermionic coordinates and the $8N^2$ bosonic momenta. We could
notate them by $|\Pi^k_\lambda \theta_\nu^a >$ where the indices mean that
the state is defined with the $(16+8)N^2$ eigenvalues. This quantum numbers
do not always completely determine the system. At least we know that when
two or more of the coordinates coincide there appear bound states at
threshold that cannot be described if we do not add another number. We shall
ignore this fact since we shall not give precise numerical results in this
section and this degeneracy would only alter the value of certain
coefficients.  The complete calculation we want to perform is

\begin{equation}
Z=\sum_{N=1}^\infty e^{-\beta \frac{N\hbar}{2R}}\int
\left(\prod_{k,\lambda}d \Pi_\lambda^k\right)
\left(\prod_{l,\mu}d \theta_\mu^l\right)
<\Pi^k_\lambda \theta_\mu^a|
e^{-\beta H\left(\Pi_\lambda^k,Y_\lambda^k,\theta_\mu^a\right)}
|\Pi^k_\lambda \theta_\mu^a>
\end{equation}

We shall first carry out the integral and then sum over the index $N$.
Let us begin with the integral over the fermions

\begin{equation}
I_f=(\hbar\alpha')^{4(N^2-1)}\int d\theta e^{-\theta^a M_a^b \theta_b}
\end{equation}
with
\begin{equation}
\{\theta^a,\theta^b\}=\delta^{a b} \sqrt{\hbar \alpha'}
\end{equation}
and where the indices shown are labels related to the gauge group as
well as to the sixteen different fermionic degrees of freedom. The constant
before the integral comes from the relation between the coordinates and
their momenta. Its function is to make the integral dimensionless.
\begin{equation}
M=i\beta\hbar^{-1}\alpha'^{-2}R\frac{1}{2\pi}Y^i_\sigma F^{\sigma\nu}_\mu 
\gamma_i
\end{equation}
 It is easy to expand the exponential in a Grassmann series. The matrix
$M$ is $16(N^2-1)\times 16(N^2-1)$ so that the $8(N^2-1)^{th}$ term is the
first to contribute. Here again, the fermions related to the identity are
not taken into account because they do not appear in the Hamiltonian. As
regards to this first term, we can proceed just ignoring all the
anti-commutators because any term coming from them and depending on
$\hbar$ cannot have all the variables we need for the integral not to
be null. Therefore, it is possible to order the operators separating
the coordinates from the momenta, and to convert it into a double
integral of Grassmann numbers, not operators. The result is known to be

\begin{eqnarray}
\int d \theta\, \frac{1}{n!}\sum_{\mbox{\tiny combinations}}
 <\theta |\prod
(-\theta_i m_{ij} \theta_j)|\theta >= \nonumber \\
=\int d \theta \frac{1}{n!}\sum_{\mbox{\tiny combinations}}
\pm \lp \prod_{i=1}^n \theta_i\rp\lp\prod m_{ij}\rp\lp\prod_{j=1}^n
\theta_j \rp= {\mbox{det}}^{1/2} M \, . 
\end{eqnarray}

The next term does not contribute because there is a repeated coordinate.
There are some terms with $8(N^2-1)+4$ operators that do contribute. Let us
show how with an example

\begin{eqnarray}
\theta_1 m_{12} \theta_2 \theta_2 m_{21} \theta_1 \left[ \cdot \cdot \cdot \right]
=m_{12}^2 \half \sqrt{\hbar\alpha'} \theta_1\theta_1 \left[ \cdot \cdot \cdot \right] = \nonumber \\
=m_{12}^2 \frac{1}{4} \hbar \alpha'  \left[
\cdot\cdot\cdot\right]. 
\end{eqnarray}
The dots inside the square brackets stand for the terms that we have already
calculated. This procedure can be generalized to give a series in $\hbar$.
Sadly, we cannot say much about the exact value of the integrals, so we
write it this way

\begin{equation} 
I_f=\sum_{k=0}^\infty \hbar^{4(N^2-1)+k} \alpha'^{4(N^2-1)+k}D^{(8N^2-8+2k)}(y)
\end{equation}
where
\begin{equation}
D^{(8N^2)}(y)=\mbox{det}^{1/2}\left(i\beta\hbar^{-1}\alpha'^{-2}R\frac{1}{2\pi}Y^i_\sigma
F^{\sigma\nu}_\mu 
\gamma_i\right) 
\end{equation}
and each $D^{(l)}$ is homogeneous of degree $l$ in the coordinates and in
general, in the elements of the matrix $M$. With this, we have exactly
calculated the classical term and obtained the dependence on the physical
magnitudes of the quantum series.

On the other hand, this is only valid for $N>1$. If $N=1$, no fermionic
operator appears in the Hamiltonian and the integral yields null. This is
to be interpreted as a consequence of the discrete nature of the spinnorial
degrees of freedom. In other words, the 'volume' occupied by the Grassmann
variables has null measure according to Berezin's rules of integration. The
appropriate count of the fermionic degrees of freedom in this case is the
sum over spin polarizations that we made for the free configurations
(classical geometry). When the Hamiltonian depends on the spin, then the
integral does contain all the information. The classical configurations are
included although they do not contribute. Both in the fermionic and the
bosonic case, they lie on spaces that have zero measure if compared to the
global phase space.

 In order to perform the bosonic integral, we expand the exponential
separating the terms in the Hamiltonian with coordinate or momentum 
dependence. Symbolically we write

\begin{equation}
e^{-\hat{H}(p)-\hat{H}(y)}=
\sum_{n=0}^\infty\frac{(-1)^n}{n!}\left[\hat{H}(p)
+\hat{H}(y)\right]^n\,.
\label{exponential}
\end{equation}
 Where $p$ and $y$ stand for any momentum or coordinate operator. We
shall keep this notation from here on. Once the binomial is expanded,
we are left with a series in the coordinates and the momenta with no
particular ordering. Taking advantage of the canonical commutation
rules we can expand each term as a finite series in $\hbar$. The
coefficients of the series will depend on the particular order of each
term and on whether the momentum and coordinate operators do have the
same indices or not. An example is

\begin{eqnarray} 
\left.\hat{p}^{2l}\hat{y}^{4 m}\right|_{\tiny disordered}
=\left.\hat{p}^{2l}\hat{y}^{4m}\right|_{\tiny ordered}+
\hbar\times \mbox{c1}\times
\left.\hat{p}^{2l-1}\hat{y}^{4m-1}\right|_{\tiny ordered}+ \nonumber \\
+\hbar^2\times \mbox{c2}\times
\left.\hat{p}^{2l-2}\hat{y}^{4m-2}\right|_{\tiny ordered}+
O\lp\hbar^3\rp
\label{ordering}
\end{eqnarray}
with $c1$ and $c2$ some constants.

The integration is carried out inserting the expansion of the
identity between the coordinate and momentum operators so as to transform
them into numbers. This insertion need just be done once in each term 
thanks to the ordering. This way we make only one integration over the 
phase space and we shall be able to extract
common properties of all terms. One of them would be integrated to give

\begin{equation}
I_b=\int dp\, dy \,\hbar^{-8N^2+2s+8N^2+k} D^{(8N^2+2k)}(y) a^s
(ap^2)^{l-s} b^{s/2}(by^4)^{m-s/2}.
\end{equation}
Where

\begin{equation}
a=\frac{\beta R}{2 \hbar}  \hspace{1cm}\mbox{and}\hspace{1cm}
b=\frac{\beta R}{16\pi^2 \hbar^2 \alpha'^3} \,.
\end{equation}

 The number of commutations that have been needed to order the operators is
$2s$. We only consider even terms because the others are related to odd
integrals that yield null.  We shall ignore again the contribution of the
identity, as a generator of $U(N)$. It can be easily calculated later. Now
we change variables $p\rightarrow a^{-1/2}p$ and $y\rightarrow b^{-1/4}y$
 and take advantage of the homogeneity properties of the $D$-functions to
simply get

\begin{equation}
I_b=\left(\frac{\beta R}{\alpha'}\right)^{\frac{3}{2}s
+\frac{3}{2}k}
\times \int dp' dy' D^{(8N^2+2k)}(y') p'^{\,2l-2s} y'^{\,4m-2s}\,. 
\end{equation}
It is manifest from here the remarkable scaling property that the
dependence on the physical magnitudes $\beta$ and $R$ is the same for
all terms in the series indexed by $l$ and $m$. This is quite fortunate
since all the bosonic integrals are, by themselves, polinomically
divergent.  We had to make this expansion  in order to get
the quantum series that disordered is hidden in the expansion of the
exponential in Eq. (\ref{exponential}). Now we can sum up the series over 
$l$ and $m$  before integrating. These are the integrals that
should be finite, at least for physical reasons, because they are the
coefficients of the quantum series.

Namely, what we wanted was to expand the original exponential of the
Hamiltonian in terms of the $\hbar$ that appears in the commutators. We know
that this is feasible because it is physically meaningful. Then we expanded
the exponential and the binomials that appeared, and after that we expanded
again each term as the series in $\hbar$ that we were looking for. As we
have found that the physical behaviour of the terms depend only on the index
$s$ of the quantum series, we are allowed to sum back again every term with
common $s$, that is, to undo the first two expansions. Therefore the
partition function for a fixed $N$ is

\begin{equation}
Z^{(N>1)}=\sum_{s=0}^\infty \sum_{k=0}^\infty g(N,s,k)
 \left(\frac{\beta
R}{\alpha'}\right)^{\frac{3}{2}s
+\frac{3}{2}k}
e^{-\frac{\beta \hbar }{2R}N}\,.
\label{par}
\end{equation}
  As we mentioned, this is not valid for $N=1$.

 The way we have made the calculation seems to appear a little clearer if we
separately compute the 'classical' $s=0$,  $k=0$ term. That amounts to ignoring
the commutation rules from the beginning or equivalently truncating the
series in Eq. (\ref{ordering}) to the first term, i.e. the $O(1)$ in
$\hbar$. That way, we can reconstruct the exponential and get

\begin{eqnarray}
Z_{\mbox{\tiny transverse}}^{\mbox{\tiny classical}}=\int dp <p|e^{-\beta
\hat{H}(p)}\mbox{det}^{1/2}
M(\hat{x})e^{-\beta \hat{H}(x)}|p>= \nonumber \\ \int dp\, dp'\, dx\,
dx'\,<p|e^{-\beta \hat{H}(p)}|p'><p'|x><x|
\mbox{det}^{1/2} M(\hat{x})e^{-\beta \hat{H}(x)}|x'><x'|p>= \nonumber\\
\int dp\, e^{-\beta \hat{H}(p)}\int dx\,\mbox{det}^{1/2} M
(\hat{x})e^{-\beta \hat{H}(x)}\,.
\end{eqnarray}
To finally arrive at

\begin{equation}
Z_{\mbox{\tiny classical}}^{(N>1)}=
\sum_{N=2}^\infty \frac{e^{-\beta\frac{N \hbar }{2R}}}{N!}
\int \left(\prod_{k,\lambda}
dy_\lambda^k\right)
\mbox{det}^{1/2} (\gamma_i y_\mu^i F^{\mu \nu}_\rho)
e^{-\frac{1}{4}\left[y^i,y^j\right]^2}\,.
\end{equation}
We explicitly show all the indices for this
 simple case and again, we have ignored the $N=1$ term. The $N!$ factor
comes again as the remnant of the gauge symmetry along the minima of the
potential. It is precisely the same result as in
Eq. (\ref{par}) and we can see that the integral that appears as a
coefficient is finite and perfectly defined. Apart from the dependence on
the temperature, the expression differs from the calculation with the free
configurations in the absence of the volume in the final result. This is due
to the fact that volume is a physical quantity that only has a complete
sense for those configurations. Here, it is taken into account in the flat,
divergent directions of the integral.  There, we would have different powers
of the volume depending on whether we integrate over the D0-brane wave
functions or their bound states. So even when we do not see powers of the
nine dimensional volume, our computation is a partition function and as such
it is not a linear function of $V_9$. The fact that these integrals include
by themselves the effects of bound states can be seen by just inserting
a $(-1)^F$ operator. This would turn this function into the Witten index,
which is precisely used to count such states. These states also include
different dependences on the volume even before perfoming the sum in $N$,
because they act as if certain directions along the classical phase space
were frozen.

Let us now complete the calculation by adding the $N=1$ term, which exactly
equals the integral over the fields related to the identity -the center of
mass of the system. This is special in several ways. Not only is it
independent of the fermionic variables, but neither does it depend on the
bosonic coordinates. Since it is just a function of the nine bosonic momenta
and the Gauss constraints are trivial, it does not receive any quantum
correction.  This is clear when we relate this special matrix configuration
with a single object in eleven dimensions: one supergraviton on its
light-cone. One single object is always free and its statistics is
irrelevant. It was already computed in the previous section, so we simply
write

\begin{equation}
Z=256\, V_9 \left(2\pi R\beta \hbar \right)^{-9/2} 
\lc e^{-\frac{\beta\hbar}{2R}} +
\sum_{N=2}^\infty\sum_{q=0}^\infty f(N,q)
 \left(\frac{\beta R}{\alpha'}\right)^{\frac{3}{2}q}
e^{-\frac{\beta \hbar }{2R}N}\rc \,.
\label{partition} 
\end{equation}
 
 We have been able to gather the two quantum series -the one of the fermions
and that of the bosons- into a single one because they are expansions in
exactly the same parameter. One of the most amazing characteristics of the
partition function here calculated is its peculiar dependence on $\hbar$. It
only appears in the center of mass contribution and in the exponential
coming from the longitudinal sum.  For all the $N>1$ terms we have
supposedly made a classical approximation followed by a quantum series that
would correct it; nevertheless the
$\hbar$'s that we include in the commutators are exactly canceled by those
that appear in the Hamiltonian. So we arrive at the bizarre conclusion that
the 'quantum' series is, in fact, not quantum at all. Moreover, this series is
an expansion in the same parameter as the series in the longitudinal
momentum $N$, therefore, we are tempted to conclude that the true natures of
both series are not clearly distinguishable. 
The function is not very different from its $\hbar\rightarrow 0$ limit, the only
role of this constant seems to be to serve as a kind of chemical potential
that measures the cost of adding longitudinal momentum to the system.

This poses the important question of how to relate this result with the
approximations we have made in previous sections and eventually, how to take
the limits that would lead us to the different string theories. Different
classical limits will not appear in this theory just as it is usual in
traditional quantum theories. According to them, specific quantum
statistical properties were effects related to the nature of the fundamental
fields. However, in Matrix Theory we have seen in a previous section that
the only way of including quantum statistics is to assume that they appear
as a consequence of the residual gauge interactions. This connects with the
surprising behaviour of the quantum series. That is why we shall see that
all these effects come out because of the appearance of new degrees of
freedom related to the off-diagonal terms. The classical limits are reached
when those degrees of freedom decouple getting infinitely massive.

Nevertheless, the decoupling of degrees of freedom is always obscure when we
look at the partition function because physical magnitudes are related to
its logarithm and derivatives thereof. Different contributions to, for
example, the Helmholtz free energy are not summed up but multiplied inside
the partition function.
 Therefore, if some of the effects are taken to be negligible, the whole
partition function may tend to zero. The best way to circumvent this
difficulty is, then, precisely, to calculate the Helmholtz
 free energy. It is, basically, the logarithm of $Z$,
which carries the advantage of turning products into sums, and so,
clarifying the discrimination of the contributions.

Let us, for example, take the classical limit
\begin{eqnarray}
Z&=&Z_{\mbox{\tiny classical}}Z_{\mbox{\tiny other}} \nonumber \\
\beta F&=&-\ln{Z_{\mbox{\tiny classical}}} -
\ln{Z_{\mbox{\tiny other}}}
\end{eqnarray}

We know that the classical partition function diverges when
$\hbar\rightarrow 0$ so that the other term must tend to zero in order for 
the complete function to remain finite. Looking at the expression for the
free energy one can see that the non-classical term acquires an infinite
energy while the classical one gets less and less energetic as we get nearer
to the limit.

It is clear that the factorization of the partition function is quite
arbitrary as we can separate any two parts by putting the appropriate
parameter and then decouple one of them by taking a certain limit for
the parameter. This is a consequence of the fact that we can organize
the degrees of freedom of the theory in many ways -with or without
physical meaning- and that, indeed, the information that the partition
function holds is global so that the behaviour of particular
configurations may be, as we have seen, quite hidden. The only physical
way of knowing which is the correct expansion in each case is to go
back to the Hamiltonian, then decide which degrees of freedom we are
interested in according to the limit we are choosing (classical, large
distances, ...), and only after that should we integrate exclusively
the appropriate configurations. Some limits will be calculated in next
section.

Let us not forget that the series is itself a limit in the sense that it is
only a good expansion for $(\beta R) \rightarrow 0$. Its validity is
basically determined by the coefficients $f(N,q)$. What we know is that the
$f(N,0)$ carry a $1/N!$ factor with more $N$ dependence coming from the
Gaussian-like integrals and that the other coefficients come from a quantum
series that should be well behaved. If we fix the radius of the eleventh
dimension, and look at the series at high temperature, we can see that the
thermodynamical behaviour is completely determined by the coefficients
$f(N,0)$ but it is clear that the quantum corrections are the only ones that
are affected by variations of the temperature.
 The question of what is the mechanism through which the theory remains
finite without any ultraviolet cut-off is unresolved yet, because it is
related to the growth of the coefficients with $N$. This would tell us how
the theory distributes its degrees of freedom. An analogous high temperature
expansion can be obtained for perturbative strings by making a high
temperature expansion for the free energy of each field in the string. This
always gives a Laurent series for $\b \rightarrow 0$ with leading term 
$\b^{-d}$ with $d-1$ the number of open spatial dimensions for the Helmholtz
free energy. The point is that the number of degrees of freedom per mass
level grows too fast so as to get the high temperature series a non
convergent one. Actually, no high temperature divergence appears because the
canonical equilibrium gets broken at the Hagedorn temperature. This is the
way modular invariance works as an ultraviolet cut-off at finite
temperature in fundamental strings.

On the other hand, for a fixed temperature, the series is not able to
describe systems with large light cone radius. It seems to be
adapted to be a good description of type IIA String Theory. Indeed,
one just has to make the substitution $R \rightarrow g_s \sqrt{\alpha'
\hbar}$ to recognize the series in $N$ as a String Theory
non-perturbative expansion with the exponential of the inverse of the
string coupling as expansion parameter. Each term is further corrected
by a weak coupling expansion, which represents the quantization around
the different solitonic sectors. This quantum series recovers its
natural parameter $\hbar$. So this would be the partition function of
type IIA String Theory including the non-perturbative effects.
Nevertheless, there is no exclusively perturbative term in this series
so that strings themselves do not seem to be accounted for. This is a
consequence of the particular choice of reference frame that we have
made in the eleven-dimensional theory which we began with. The length $R$ we
have been writing is not exactly the $R^{11}$ that is what is directly related to the
string coupling but the same radius measured by a strongly boosted
observer. The same happens to the temperature, it is an energy and so
it is affected by Lorentz transformations. Anyway, this is not much
worrying since we have calculated a scalar dimensionless
quantity; that is, the partition function is the same no matter if it
is calculated in the light-cone or in any other reference frame. This
way, we can invert the boost and make the substitution
\begin{equation}
\left.\frac{\beta R}{\alpha'}\right|_{\mbox{\tiny{light-cone}}} \rightarrow
\left.\frac{\beta R^{11}}{\alpha'}\right|_{\mbox{\tiny{at rest}}}\simeq
\left.\frac{\beta R}{\alpha'}\right|_{\mbox{\tiny{light-cone}}}
\times\sqrt{2}\lp \frac{R^{11}}{R}\rp^2 \,.
\end{equation}
with
\begin{equation} 
\frac{R^{11}}{R} \rightarrow \infty
\end{equation}  
It may be useful to remind that when we say 'at rest' we mean that the
observer measures the invariant radius of the eleventh dimension, the
smallest, so that, somehow, he is at rest with respect to the compact
circle. In this reference frame, the parameter is much bigger
so that the series is useless except for extremely high temperatures.
Therefore, in spite of being Lorentz invariant, the series is
only adapted to describe light-cone objects. 
Now the relation to the type IIA string is more direct but
still, we do not seem to have any string! The reason for this is that
our light-cone calculation has 'integrated out' the string degrees of
freedom and spreaded them over the D0-brane ones. The degeneracies that
are counted by the series in the index $N$ do not either come from 
strings nor D0-branes but rather from what in Solid State Physics
would be called $quasi$-D0-$branes$.
These are the original branes, but dressed by the closed type IIA strings,
other D-branes and anti-D-branes and
interacting through them. Those are the solitons to which the series
refers and this is the reason why we do not have any
zero mode. The conclusion is that our calculation corresponds to the
partition function of type IIA strings, but it is adapted to a point of
view different from the usual one.

\section{The field-theoretical approach}

In the previous section we have computed the partition function for the
Matrix Model using the  commutation properties of the matrices that describe the
D0-brane dynamics. This section is devoted to a calculation of $Z(\beta)$ 
using the perturbative expansion for the $U(N)$ Super Yang-Mills Quantum
Mechanics. This type of calculations are closely related to those done in
\cite{bb,danfer}, where the one loop effective potential between two D0-branes
is obtained. We will also compute the thermal degrees of freedom for the
case of the Matrix Model on $T^2$, using the dual description in terms of 
the D-strings dynamics. Comparing both analysis we will check $T$-Duality.

It is assumed that, when studying the interaction of non-BPS states, the
correspondence between Matrix Model and Supergravity calculations may be
realized at finite temperature \cite{malda,malstr,kle}. This picture comes
from the idea that the non-extremality of the D-branes would be included
into an entropy, and it is closely related to the Supersymmetry breaking
produced by the different statistics of fermionic and bosonic
fields\footnote{In a recent paper \cite{tse} closely related to this part of
our work, A. Tseytlin has analyzed these ideas.}.

We will work with the  Euclidean version of the SYM action, that is (in the
temporal gauge, $A_0=0$)
\begin{equation}
S^E=\int_0^{\beta}d\tau\, {\mbox {Tr}}\left[
\frac{\dot{Y}^i\dot{Y}^i}{2R}+\frac{1}{\sqrt{\alpha'}} \theta^T \dot{\theta}+
\frac{R}{16 \pi^2 \alpha'^3}\left[Y^i,Y^j\right]^2+\frac{R}{2 \pi
\alpha'^2}
\theta^T\gamma^j\left[Y^j,\theta\right]\right] 
\label{action}
\end{equation}
where the Euclidean time $\tau$ is taken to be a circle of length
$\beta$ and we have set $\hbar = 1$ again. The $Y^i$ are nine scalars
fields and the $\theta$ are Grassmann fields of the transverse $SO(9)$
rotation group. Both are matrices of the adjoint of $U(N)$.

We will compute the one-loop correction to the free partition function for
the fields that appear in the Matrix Lagrangian, and compare to the
results obtained in  the previous sections. The physical meaning of
this type of calculations will be explained in detail below. We will show
how this description, like the calculation presented in \cite{greenmavaz},
 only takes into account the thermal properties of the fields attached to
the D0-branes without thermalizing the D-particle itself.

As usual in quantum field theory, we can obtain  the corrections to the partition
function by using the Feynman diagram techniques. Expanding the
interaction term of the action in the Path Integral functional we have
\begin{equation}
Z(\beta,N)=Z_0(\beta)\left[1+\sum_{n=1}^{\infty}\frac{(-1)^n}{n!}{\cal
Z}_n(\beta)\right]\,, 
\label{vertex}
\end{equation}
where ${\cal Z}_n$ stands for the $n$-vertex correlation functions. As an
example, in a simple model with two Yukawa-coupled real fields we would have 
\begin{equation}
{\cal Z}_n=\int \prod_{i,j,k}^{n} dx_i dy_j dz_k \int{\cal D}\phi{\cal D}\psi 
\left(\lambda \phi(x_i)\psi(y_j)\phi(z_k)\right)^n
e^{-\left(S^E_{free}\left[\phi\right]+S^E_{free}\left[\psi\right]\right)}
\label{correlation}
\end{equation}
and
\begin{equation}
Z_0(\beta)=\left[\mbox{det}\left(K(x,\beta,\phi)\right)\right]^
{\frac{(-1)^F}{2}}\left[\mbox{det}\left(K(x,\beta,\psi)\right)\right]^
{\frac{(-1)^F}{2}}
\end{equation}
with $F$ the fermionic number. The above  expressions may be  trivially 
generalized to the momentum space. 

We will start by computing and studying the tree level contribution to the 
Helmholtz free energy by using the Schwinger representation of the field 
propagator. Following this method we write 
\begin{equation}
-\beta
F_0(\beta)=-\mbox{Tr}\left[(-1)^{F}\int_0^{\infty}\frac{dt}{t}e^{-\left(k^2+m^2)\right)
t}
\right]. 
\end{equation}
In our case we have massless fields in $d=0+1$ dimensions, then  the
trace in the above expression has to be taken over the Matsubara
frequencies and over internal degrees of freedom which are the
representations of the transverse $SO(9)$ and the $U(N)$ gauge group.
As the result of a blind calculation, we would propose
\begin{equation}
-\beta F_0(\beta)\stackrel{?}{=}-\frac{N^2}{2}\int_0^{\infty}dt
\,t^{-1}\left [9\,\theta_3\left(0,\frac{2 \pi i
t}{\beta^2}\right)-8\,\theta_2\left(0,\frac{2 \pi i
t}{\beta^2}\right)\right].
\label{free}
\end{equation}
Here the $N^2$ factor comes from the $U(N)$ degrees of freedom, and the
multiplicative factors of the thermal modular functions reflect the number
of bosonic $(9)$ and fermionic $(8)$ fields.
Some comments about the previous expression are needed. If we were studying
a second quantized theory of objects living in eight dimensions we would 
expect the Helmholtz free energy to grow proportionally to the eight-dimensional
volume. As we see $F_0(\beta)$ in (\ref{free}) is volume independent. This
fact shows that what we are really doing is to thermalize the internal
degrees of freedom of the D0-branes, that is, the strings attatched to them.

This type of situation has already  been studied from the string point of view.
However it seems that there are some subtleties to be taken into account to
establish the possible connections. In \cite{greenmavaz} the thermal free
energy of an open superstring gas in presence of a Dp-brane has been
computed, obtaining for the special $p=0$ case
\begin{equation}
-\beta F^{\mbox{{\tiny String}}}(\beta)=-\frac{N^2}{8}\int_0^{\infty}\frac{dt}{2}
\,t^{-1}\left [\theta_3\left(0,\frac{2 \pi i
t}{\beta^2}\right)-\theta_2\left(0,\frac{2 \pi i
t}{\beta^2}\right)\right]f(t)  
\label{string}
\end{equation}
where $f(t)$ is the  partition function of the Type I string theory.
If one takes the zero temperature limit one recovers the vanishing vacuum
energy of a supersymmetric theory. This property have been used by Polchinsky
\cite{pol} to show the BPS nature of the D-branes. If we want to compare this
expression with (\ref{free}) we must  consider only the massless field
content of the string spectrum giving
\begin{eqnarray}
-\beta F^{m=0}(\beta) &=& - \frac{N^2}{2}\int_0^{\infty}dt
\,t^{-1}\left [8\,\theta_3\left(0,\frac{2 \pi i
t}{\beta^2}\right)-8\,\theta_2\left(0,\frac{2 \pi i
t}{\beta^2}\right)\right]= \nonumber \\
=&-& \frac{8\,N^2}{2}\int_0^{\infty}dt
\,t^{-1}\theta_4\left(0,\frac{2 \pi it}
{\beta^2}\right).
\label{massless}
\end{eqnarray}
Where the $8$ corresponds to the number of degrees of freedom of the
massless field. Remember that at this level the string spectrum is composed
by a vector and a Majorana-Weyl fermion. Both belong in their respective
$SO(8)$ representation. However in (\ref{free}) we would have obtained
a different result that could be written 
\begin{equation}
-\beta F_0(\beta)\stackrel{?}{=}-\beta F^{m=0}(\beta)- \frac{N^2}{2}\int_0^{\infty}dt
\,t^{-1}\theta_3\left(0,\frac{2 \pi it}
{\beta^2}\right).
\end{equation}
The last term in the previous equation prevents $F_0(\beta)$ to vanish
at $T=0$ because of the $\theta_3(0,s)$ zero mode \footnote{If we
neglect the zero mode of the fields, we see that both expressions
vanish in the zero temperature limit, but they will do it in a
different way.}. This property of the Helmholtz free energy we have
obtained for the Matrix Lagrangian seems to break the supersymmetric
nature of the theory in such a way that the D-particles would become
non-BPS states, but there are some details related to the gauge we have
chosen that will restore our standard knowledge about D-branes.

In the string calculation, the Dp-brane configuration is obtained by
compactifying Type I string theory on a torus with radii
$R_{p+1},...,R_{d-1}$ which then are taken to zero. The free energy is then
calculated integrating the momentum of the string coordinates with Neumann
boundary conditions. When we do it we fix the D-brane at a given
position in the Dirichlet direction; only allowing fluctuations on the
D-brane world-volume. In other words this type of computation only takes  
into account the string's knowledge of the full $d$-dimensional space-time,
that gives the eight bosonic and fermionic contributions in (\ref{string}). 
In the particular case of a D-particle we assume it to be fixed at a given 
point in space, and we compute the string contributions to $F(\beta)$.
From the point of view of the SYM Quantum Mechanics that describes the
D-particle dynamics, and then from the Matrix Model one, this kind of
configuration must correspond to an election of a fixed background. However
this is not enough. If we let all  the D-particle coordinates fluctuate we
would once more have nine bosonic contributions without connection to the
string calculation. What we are forced to do is to assume that there is one
non-fluctuating direction. To choose the frozen direction we can take
advantage of the worldsheet conformal invariance of string theory that forbids
the vibrations along the longitudinal direction of the string. In our case this
coordinate coincides with the straight line that connects the D-particles
\cite{danfer,rutgers}. Consequently, if we want to relate the Matrix Model 
computation and its string origin we have to freeze this direction. 

We can analyze this problem from a more technical perspective. As we said 
the theory that describe the D-particle dynamics is a SYM quantum
mechanics coming from a dimensional reduction of the theory in  $d=9+1$ 
to $d=0+1$. We can start our analysis taking the Arnowitt-Fickler gauge 
in the initial SYM quantum field theory, that is
\be
n_{\mu}A^{b}_{\mu}=0 
\label{arno}
\ee
where $n_{\mu}$ are components of a unitary vector. Here the gauge fields
are functions of the full ten-dimensional space, then after the dimensional
reduction we have
\be
n_{0}A^{b}_{0}+n_{j}Y^{b}_{j}=0 \hspace{1cm} j=1,...,9\, .
\label{arnoredu}
\ee
Remember that the condition in (\ref{arno}) does not totally reduce the 
degrees of freedom to the physical ones. So  we must take another gauge
condition to complete the reduction. Here we can choose the temporal gauge 
$(A_0^{b}=0)$, that reduces the relation in (\ref{arnoredu}) to a constraint 
between the scalars fields $Y_j^b$. This picture is what underlies the
physical idea of freezing the vibrational modes of the string along the
longitudinal direction, and corresponds to the picture described in the
previous section for the hamiltonian formalism.
  
We could have done the same analysis by choosing another gauge and
introducing the adequate ghost contribution. As in \cite{bb} it is possible
to choose the background field gauge, in the special case of the coinciding
D-brane configuration. In this case the spectrum of the theory is composed
of eight on-shell bosons, after considering the ghost contribution, eight
fermion fields, and the interactions between them are the same we have in
the temporal gauge.

To summarize, we have then shown that, for the  $d=0+1$ gauge theory
describing the dynamics of our system, the temporal gauge does not
completely reduce the degrees of freedom to the physical ones. Another
reduction is then needed. This constraint has to be taken from the
dimensionally-reduced theory we started from. In the case of the D-particle
dynamics it is its string vibration origin what determines the reduction.

Let us come back to the volume dependence of (\ref{massless}). From an
exact M-theoretic point of view we should compute the contribution coming
from the nine directions. In fact we would expect a linear volume dependence
of the Helmholtz free energy. These contributions could be studied by
assuming that we are separating the physical degrees of freedom by
means of a Born-Oppenheimer approximation \cite{danfer}. We are
assuming that the phase space of the theory may be factorized in
terms of the dynamics of the D-particles as objects moving in the nine
transverse directions and their vibrational degrees of freedom. The
eight transverse string modes are decoupled from the motion of the
D0-brane which must be studied as we did in the previous section.
Finally we may describe the full dynamics in terms of a phase space
coming from the motion along the nine dimensions and characterized by
the presence of the string vibrational degrees of freedom for each
point in that space. This idea have been also proposed in relation with
$p$-brane and black-hole physics \cite{blackholes1,tse,kle}.
 
Let us start a more detailed study  assuming that the $N$ D0-branes are 
located in a background characterized by
\begin{equation}
Y_1=\left(\begin{array}{cccc}
x_1   & 0      &  0  & \cdots \\
0     & x_1    &  0  & \cdots \\
\vdots& 0      &  x_1& \cdots \\
\vdots& \vdots &  0  & \ddots \\
0     &   0    &  0  & 0    \\
\end{array}
\right).
\label{bakg}
\end{equation}  
This election corresponds to a fluctuation of the  D-particles orthogonal to the
$Y_1=x_1$ fixed plane. This configuration completely  decouples the $Y_1$
field from the action. On the other hand we  consider the  transverse 
vibration of the massless fields connecting the branes. We should do that  
taking the following parameterization

\begin{equation}
\left(Y_k\right)^i_j=x_k\delta^i_j+\lambda\phi^i_j
\end{equation}
with $\phi^i_j \in U(N)$. This corresponds to coinciding D0-brane
positions,  preserving the full $U(N)$ gauge symmetry. This
configuration exactly coincides with the string approach to the thermal
strings stretched between $N$ D-branes in \cite{greenmavaz}.

It is possible to generalize our study to the separated branes
configurations. As it is well known, separating the D-branes breaks the
$U(N)$ symmetry down to $U(N_1) \times U(N_2)$. It is easy to see how this
breaking occurs. As an example we can look at the $N=2$ case. Taking the
branes in different positions in the frozen direction, we have
\begin{equation}
Y_1=\left(\begin{array}{cc}
x_1   & 0  \\
0     & x_2 \\
\end{array}
\right)
\,\,\mbox{and}\,\,
Y_i=\left(\phi_i^a T^a
\right)
\end{equation}
where $T^a$ are the group generators. One can easily obtain the mass correction 
to the scalar fields, coming from the interaction term in the action
which is
\begin{equation}
\mbox{Tr}\left[Y_1,Y_i\right]^2=\phi_i^1\phi_i^1\frac{(x_1-x_2)^2}{4}+
\phi_i^2\phi_i^2\frac{(x_1-x_2)^2}{4}
\end{equation}
that is diagonalized in terms of the massive fields $\phi_i^1$ and $\phi_i^2$.
This property was firstly pointed out by E. Witten in \cite{bound}.
Finally the free energy reads

\begin{equation}
-\beta F(\beta)=-\frac{8}{2}\sum_{i=1}^{2}\,N_i^2\,\int_0^{\infty}dt
\,t^{-1}e^{-\left(x_1-x_2\right)_i^2}\theta_4\left(0,\frac{2 \pi i
t}{\beta^2}\right).  
\label{N2}
\end{equation}
The same arguments hold for an arbitrary $N$ case. If we separate this system 
into two objects with $N_1$ and $N_2$, the free energy will have the
form 
\begin{equation}
-\beta F(\beta)\simeq -\frac{N_1^2}{2} G_1(\beta)-\frac{N_2^2}{2}G_2(\beta). 
\end{equation}
Where $G_1(\beta)$ and $G_2(\beta)$ are the functions of the temperature
corresponding to the thermal degrees of freedom of the fields living
attached to
$N_i$ block of D-particles. Following this procedure it is possible
to change the $N^2$ dependence in (\ref{massless}) into a linear one on $N$,
that corresponds to a $U(1)^N$ gauge group. The description we have shown in
terms of the string physics included in the D0-brane dynamics may be done
without any reference to it.  As we have said for the overlapping D-brane
configuration, here we may also fix any other gauge condition and obtain the
same results as with the temporal (Arnowitt-Fickler) gauge. The spectrum
obtained pulling apart the particle positions is, as expected, gauge
independent. It is worth noticing that, as in quantum field and string
theories \cite{moore}, the interaction can be partially absorbed as a
correction for the masses of the fields.

Now let us  carry out the computation of the one-loop correction to
the free energy. As we did  at tree level we have to decide how many string
directions we take into account. Assuming the philosophy of the
Born-Oppenheimer approximation we have explained before, we will fix the 
corresponding longitudinal direction. Following \cite{bb,compact} we rewrite the 
action (\ref{action}) in the usual SYM form, which reads
\begin{equation}
S^E=\frac{1}{g^2}\int_0^{\beta}d\tau Tr\left[\dot{Y}^i\dot{Y}^i+\theta^T
\dot{\theta}+\frac{1}{2}\left[Y^i,Y^j\right]^2+
\theta^T\gamma^j\left[Y^i,\theta\right]\right]
\label{SYM}
\end{equation}
with
\begin{equation}
g=\frac{R^{3/2}}{2\pi\alpha'^{3/2}}
\end{equation}
An adequate rescaling of the  bosonic and fermionic fields with the SYM coupling 
$g$ allows us to study the Feynman rules of this theory. It is also  useful to
explicitly write down the matrix fields in terms of the U(N) generators,
obtaining\footnote{Here we take the normalization
$\mbox{Tr}\lp T^a T^b\rp =\delta^{ab}$}
\begin{equation}
S^E=\int_0^{\beta}d\tau \left[\dot{Y}_a^i\dot{Y}_a^i+\theta_a^T
\dot{\theta}_a+\frac{g^2}{4}\,Y_a^iY_b^jY_f^iY_g^j F^{abc}F^{fgc}+       
g\,\theta_b^T\gamma^jY_c^i\theta_d F^{cdb}\right].                   
\label{component}   
\end{equation}
To be complete  we can write a formal expression for the
$n$-vertex correction to the Helmholtz free energy, which is 
\be
F(\beta)=- \frac{1}{\beta}\log Z_0(\beta)-\frac{1}{\beta}\log 
\left[1+
\sum_{n=0}^{\infty}\frac{{\cal Z}_n(\beta)}{n!}\right].
\ee
We show how to compute the functions ${\cal Z}_n$ for a very simple case in
(\ref{correlation}). In the case of the Matrix Model  we shall write 
${\cal Z}_n(\beta)$ as
\be
{\cal Z}_n=\int\,{\cal D}Y^i_a\,{\cal D} \theta_b
\,e^{-\left[S_Y^{Free}+S_{\theta}^{Free}\right]}\sum_{k+p=n}\frac{(-1)^k}{k!}
\frac{(-1)^{2p}}{p!}
\left[I(Y^i_a)\right]^k\left[I(\theta_a,Y^i_a)\right]^{2p}
\label{z_n}
\ee
where $I(Y^i_a)$ and $I(\theta_a,Y^i_a)$ are the interaction terms appearing
in the Matrix Lagrangian.

\begin{figure}
\let\picnaturalsize=N
\def\picsize{2in}
\def\picfilename{diagra.ps}
\ifx\nopictures Y\else{\ifx\epsfloaded Y\else\input epsf \fi
\let\epsfloaded=Y
\centerline{\ifx\picnaturalsize N\epsfxsize \picsize\fi
\epsfbox{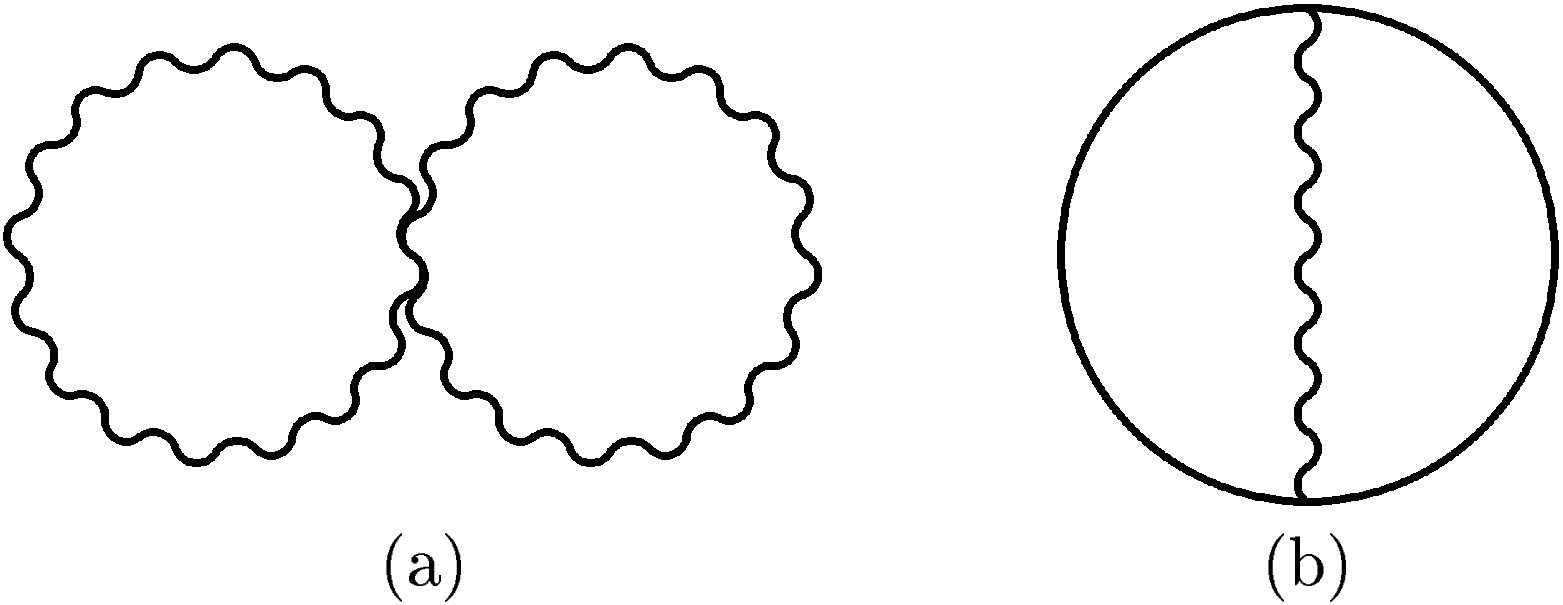}}}\fi
\caption{\small {The one-loop Feynman diagrams involving scalar fields
(a) and the coupling of fermions and scalar fields (b)}}
\label{diagramas}
\end{figure}

Now we can explicitly write the Feynman rules of the Lagrangian in
(\ref{component})
\begin{eqnarray}
\mbox{Scalar propagator}&=& \frac{\delta^{ab}\delta_{ij}}{p^2}
\hspace{1cm} 
\mbox{with} \, p=p_0=\frac{2 \pi n}{\beta} \,(n \in Z) \nonumber \\
\mbox{Fermion propagator} &=&- \frac{i \delta^{ab}}{\gamma^k p_k}\hspace{1cm}
\mbox{with} \, p=p_0=\frac{(2n+1)\pi}{\beta}\,(n \in Z)\nonumber
\end{eqnarray}
and the interaction vertices are
\begin{eqnarray}
&\mbox{i)}&\,\frac{g^2}{4}Y_a^iY_b^jY_f^kY_g^l F^{abc}
F^{fgc}\rightarrow     -{g^2}\left[F^{abc}
F^{fgc}\left(\delta_{ij}\delta_{kl}-\delta_{kj}\delta_{il}\right)+\mbox{
permutations}\right]\nonumber\\
&\mbox{ii)}&\,g\theta_b^T\gamma^jY_c^j\theta_d
F^{cdb}\rightarrow-igF^{cdb}\gamma^j
\label{rules1}
\end{eqnarray}
in both cases the total momentum of the incoming fields is zero. It is easy to
check that the correlation functionals in (\ref{correlation}) vanish for
an odd number of fermionic fields, leaving the perturbative expansion of
$Z(\beta)$ to be a power series in $g^2$. Finally we must take care of
the discrete nature of the  momenta when we compute the loop integrals. 
The one-loop correction we want to obtain corresponds to the diagrams
in Fig. \ref{diagramas}.

The purely scalar contribution is
\begin{eqnarray}
{\cal Z}_1^{scalars}(\beta) & = & - 336\,g^2 F^{ade}F^{ade}\frac{\beta^3}{\left(2
\pi\right)^3}\sum_{n,m}\frac{1}{n^2}\frac{1}{m^2}\nonumber \\
& = & -336\, g^2\,\left[\frac{(N^2-1)(N^2-2)}{2}\right]\frac{\beta^3}{\left(2
\pi\right)^3}\sum_{n,m}\frac{1}{n^2}\frac{1}{m^2}.
\label{bosons} 
\end{eqnarray}
The factor $336$ appears because of the $8$   bosonic oscillating
directions, i.e.  after  freezing the longitudinal one.

The contribution coming from the diagram involving fermionic fields gives
\begin{eqnarray}
{\cal Z}_2^{fermi}(\beta) = -2\,g^2 \gamma_j\gamma_j F^{fae}F^{fae}
\,\frac{\beta^3}{\pi^3}\sum_{n,m}\frac{1}{2n+1}\frac{1}{2m+1}
\frac{1}{ \left(n-m \right)^2} 
\nonumber 
\\
= - 256\, g^2 \left[\frac{(N^2-1)(N^2-2)}{2}\right]\frac{\beta^3}
{\pi^3} \sum_{n,m}\frac{1}{2n+1}\frac{1}{2m+1}
\frac{1}{\left(n-m\right)^2}
\label{fermi}
\end{eqnarray}
where we have taken the trace over the fermionic degrees of freedom included
into the $\gamma_j$. We are now able to study some properties of the
one-loop correction to the partition function, which takes the form
\begin{equation}
{\cal Z}_1(\beta)=-{\cal Z}_1^{scalars}(\beta)+
\frac{1}{2}{\cal Z}_2^{fermi}(\beta).
\label{z1}
\end{equation}
Then the corresponding free energy looks like
\begin{eqnarray}
F(\beta)& = &- \frac{1}{\beta}\log Z_0(\beta)-\frac{1}{\beta}\log 
\left[1+{\cal Z}_1(\beta)
\right]=\nonumber \\
& = & F_0(\beta)-\frac{1}{\beta}\log\left[1+g^2
\left[\frac{(N^2-1)(N^2-2)}{2}\right]\frac{\beta^3}{\pi^3} \,G\right]
\label{loop}
\end{eqnarray} 
where $G$ is a constant coming from (\ref{z1}). The value of this constant
can be obtained by a generalized-$\zeta$ function regularization. In the
zero-temperature case this contribution vanishes by dimensional
regularization because of the fact that there is no dimensionful
parameter involved. This means that we are dealing with a high temperature
expansion whose relationship to the $T=0$ case is subtle. 

We can now check what happens if we consider the Matrix Model compactified
on a torus. The dynamics of $N$ D0-branes on  $T^n$ is described by a system of 
$N$ D$n$-branes on the dual torus \cite{bfss,intro,compact,taylor}. In our
case we start with the Matrix Model on $S^1(R) \times (S^1(L) \times {\bb
R}^8)$, where the first $S^1$ stands for the compactified eleventh
dimension. In this case the dual description is given by a system of
D-strings wrapped around a cylinder of length $\Sigma=1/L$. The theory that
describes this system is a $d=1+1$ SYM quantum field theory coming from
dimensional reduction from $d=9+1$.  Because of the appearance of the
$\Sigma=1/L$ multiplicative factor in the dual action \cite{taylor} we can
redefine the coupling constant of the theory in such a way we can write
\begin{equation}
g^2_{dual}=\tilde{g}^2=g^2 \Sigma
\end{equation}   
where $\tilde{g}$ is the effective coupling constant in the Kaluza-Klein
sense.

In this case the gauge condition in (\ref{arno}) reduces to the following
form
\be
n_0A^{b}_0+n_1A^{b}_1+n_jY^{b}_j=0
\ee
then the  problem of gauge fixing in this case is exactly the same as
the one  we have analyzed before. After taking the temporal gauge  
we have to choose the additional
condition coming from the residual constraint that relates the eight scalar
fields $Y_j$ and the $A_1$ gauge field. Initially we have two possible
options, either we can set the gauge field to zero, assuming that there are
 no propagating modes on the D-string, or we can stop the string vibration in one
direction. In the latter case  we will have a purely Neumann string mode propagating 
on the D-string and interacting with seven scalars. 
In both cases we will have eight bosonic and fermionic degrees of freedom. 
At the tree level the free energy is the same as in (\ref{massless}), 
except for the Kaluza-Klein modes contribution
\begin{eqnarray}
-\beta F^{\Sigma}(\beta) &=& - \frac{N^2}{2}\int_0^{\infty}dt
\,t^{-1}\left [8\,\theta_3\left(0,\frac{2 \pi i
t}{\beta^2}\right)-8\,\theta_2\left(0,\frac{2 \pi i
t}{\beta^2}\right)\right]\theta_3\left(0,\frac{2\pi it}{\Sigma^2}\right)=
\nonumber \\
=&-& \frac{8\,N^2}{2}\int_0^{\infty}dt
\,t^{-1}\theta_4\left(0,\frac{2 \pi it}
{\beta^2}\right)\theta_3\left(0,\frac{2\pi it}{\Sigma^2}\right).
\label{torus}
\end{eqnarray}
As we can easily see, we recover the free decompactified limit, that is the
Matrix Model result, when we take the dual radius $\Sigma$ to zero.

It is now easy to compute the one-loop corrections to the Helmholtz free energy 
for the compactified case. The interaction terms we have are independent of
the gauge reduction and exactly map into the open Matrix-Model ones.
We obtain
\ber
{\cal Z}_1^{scalars}(\beta) =  - 336\, F^{ade}F^{ade}
\frac{\tilde{g}^2}{\Sigma}
\frac{\beta^3}{\left(2\pi\right)^3}
\sum_{n,m}\sum_{l,s}
\frac{1}{\left[n^2+\left(m\frac{\beta}{\Sigma}\right)^2\right]}
\frac{1}{\left[l^2+\left(s\frac{\beta}{\Sigma}\right)^2\right]}
\nonumber \\
 = -336\,\left[\frac{(N^2-1)(N^2-2)}{2}\right]
\frac{\tilde{g}^2}{\Sigma}
\frac{\beta^3}{\left(2\pi\right)^3}
\sum_{n,m}\sum_{l,s}
\frac{1}{\left[n^2+\left(m\frac{\beta}{\Sigma}\right)^2\right]}
\frac{1}{\left[l^2+\left(s\frac{\beta}{\Sigma}\right)^2\right]}.
\label{torusbos} 
\eer
To check $T$-duality in this expression, we have to rewrite the dimensionless 
coupling $\tilde{g}$ in terms of the initial Matrix parameter, the
coupling  constant $g$ and finally take the limit $\Sigma
\longrightarrow 0$, which reduces this  to (\ref{bosons}). The same arguments hold
for the fermionic contribution (\ref{fermi}). We can give a general argument
for the $n$-vertex function in (\ref{z_n}). In this case, taking
the limit of the open Matrix Model (that corresponds to the
dimensionally reduced  D-string system), one gets
\be
{\cal
Z}_n\propto\left(\frac{\tilde{g}^{2}\beta^{3}}{\Sigma}\right)^n=
(2 \pi)^{-2n}\left(\frac{\beta R}{\alpha'}\right)^{3n}.
\ee
This parameter is explicitly independent of the compactification radius,
therefore it is self-dual and then coincident with that of the already open
Matrix Model.

The expression in (\ref{z_n}) corresponds to the term with $g^{2n}$
(or $\tilde{g}^{2n}$ , when we compactify one dimension) in 
the perturbative expansion of the free energy 
\be
F^{\Sigma}(\beta) = F^{\Sigma}_0(\beta)-\frac{1}{\beta}
\log\left[1+
\sum_{n=0}^{\infty}\frac{\tilde{g}^{2n}\beta^{3n}}{\Sigma^n}
G_n\left(\frac{\beta}{\Sigma}\right)\right]
\label{looptorus}
\ee 
where $G_n$ is the function that comes from the diagram of order 
$\tilde{g}^{2n}$. The regularization arguments given for (\ref{loop}) hold here
too. This function depends on the Kaluza-Klein modes of the compactified
dimension but, in the small $\Sigma$ limit, it only takes into account the
corresponding zero mode. The behaviour of the free energy of a D-string
system ( Eq.(\ref{looptorus})) allows us to explicitly check $T$-duality
order by order in perturbation theory. In fact if we computed a general
$n$-loop contribution we would see that its dependence of the coupling
constant, the temperature, and the length of the compact dimension would map,
after taking the corresponding limit, into the $d=0+1$ SYM computation of
the same contribution. This property is not surprising because of the string
origin of the $T$-duality of the D-brane dynamics.

We can now go back to the physics involved in the Born-Oppenheimer
approximation. The complete partition function of the finite-N Matrix model
has to be obtained multiplying the internal (stringy) degrees of freedom and
those of the translational D-particle dynamics. In terms of phase spaces we
can express the complete phase space as the direct product of the
translational and internal ones. This assumption amounts to the following
form for the partition function
\be
Z^{\mbox{{\tiny Born-Oppenheimer}}}(\beta)= Z^{\mbox{{\tiny
Free}}}(\beta)\times Z^{\mbox{{\tiny Internal}}}(\beta).
\ee 
Finally, let us mention that adding the internal partition function to one
apropiate term in the sum in (\ref{object}) we recover the desired
$N^2$ dependence coming from the internal unbroken $U(N)$ SYM degrees of
freedom which are necessary to fit the entropy of a given $N$-parton in
(\ref{object}) as that of a Schwarzschild black hole of the model in 
\cite{blackholes1}, and obtain the Bekenstein-Hawking area law.

\section{Conclusions}

  Matrix theory shows several different aspects when one looks at its
thermodynamical behaviour. When the distances between objects are large,
they can be described by a semiclassical expansion around what can be later
interpreted as the Kaluza-Klein modes of a supergraviton in eleven
dimensions. This is, in fact, the low energy limit of M-theory and it is the
only case in which we have been able to take the radius of the eleventh
dimension  to infinity and see what
are the predictions of the model for a completely open universe. We have
seen that, as maybe one could have expected from the beginning, it
precisely corresponds to the results obtained for  eleven-dimensional
supergravity, defined in an open space.

One interesting point would be the connection between our work and the fact
 that the semiclassical Born-Oppenheimer
approximation for the supergraviton modes justifies their election in
\cite{blackholes1} as candidates to describe black holes.  The
contribution of their center of mass does not spoil the arguments
exclusively based on the Super Yang Mills entropy. Another point of view is
taken in \cite{blackholes2}, where the model supposes that a black hole
consists of a gas of distinguishable D-particles with Boltzmann statistics.
In this case, the separation between fast and slow modes of the system helps
us justify the dependence of the entropy with the number of D-particles.
Indeed, the field theory gauge symmetry is broken to $U(1)^N$ and it seems
that $S \propto N$ may not be taken as an assumption. It keeps being unclear
how the discernibility is acquired in this model. Maybe one could get some
insight about this possible connection from the calculation in section four,
where the whole gauge group is included. The question is that we do
not really see how the black hole enters into the stage in our picture.

   In section four we have tried to calculate some global aspects of the
partition function including interactions and quantum effects. We have seen
that the consequence of taking canonical Heisenberg commutators between the
coordinate fields and their conjugate momenta is the appearance of a  series
whose natural parameter is not $\hbar$ as usual but the light-cone radius.
In fact, the parameter is the same as that of the series coming
from the sum over the possible values of $p^+$. We see this as
a confirmation of the idea that this theory is quantum in a broader
sense that the presently known quantum theories and that String
Theories and Supergravities come out as classical limits of a moduli space.

  It is known, both in field and string theory, that the corrections to the
free theory coming from interactions can always be seen as the appearance of
effective masses that correct the original ones. We wanted to see that in
our case, but to do it, we needed to make a Born-Oppenheimer approximation.
It consists in separating the 'movements' of the system in fast and slow
ones. That is what we made in section 5. We suppose that we can divide the
system in clusters of supergravitons with weak interactions among them. Each
of these objects has low energy so we can separate the fields
that propagate along their  world-line (fast movements) and
the  displacements of their centers of mass through the target space (slow
ones). The world-line calculation has been performed
and represents a first attempt to include interactions as well as quantum
effects that do not have statistical nature. We have obtained a high
temperature expansion that corresponds to the series in the Super Yang-Mills
coupling constant. This calculation and the equivalent one using T-duality
and D-strings can be used to relate this approximation with one loop
expansions in string theory like those made in \cite{greenmavaz}.
They can also serve as a finite temperature check of T-duality, that
holds to all orders in perturbation theory, as expected.

\newpage

\section{Acknowledgments}
We thank J. L. F. Barb\'{o}n, J. Casahorr\'{a}n, J. Gaite and J. Ferrer 
for useful discussions.


\begin{thebibliography}{99}

\bibitem{bfss}  T. Banks, W. Fischler, S. H. Shenker and L. Susskind,
"M Theory As A Matrix Model: A Conjecture",{\em Phys. Rev.}{\bf D55} (1997)
5112-5128,  hep-th/9610043.

\bibitem{intro} A. Bilal, "M(atrix) Theory: a Pedagogical Introduction",
LPTENS-97/43, hep-th/9710136
\\
T Banks,"Matrix Theory",hep-th/9710231
\\   
 D. Bigatti, L. Susskind, "Review of Matrix Theory",hep-th/9712072.

\bibitem{dlcq} L. Susskind, "Another Conjecture about M(atrix) Theory",
SU-ITP-97-11 and hep-th/9704080
\\
N. Seiberg, "Why is the Matrix Model Correct?",
{\em Phys. Rev. Lett.}{\bf 79} (1997) 3577-3580,
hep-th/9710009
\\
S. Hellerman and J. Polchinski, "Compactification in the
Lightlike Limit", NSF-ITP-97-139 and  hep-th/9711037
\\
D. Bigatti, L. Susskind,"A note on discrete light cone
quantization",hep-th/9711063.


\bibitem{compact}W. Fischler, E. Halyo, A. Rajaraman and L. Susskind, "The
Incredible Shrinking Torus", {\em Nucl. Phys.}{\bf B501}
 (1997) 409-426, hep-th/9703102 \\
A. Sen, "D0 branes on $T^n$ and Matrix Theory",
MRI-PHY/P9709
and hep-th/9709220.

\bibitem{us}  M. Laucelli Meana, M. A. R. Osorio, and J. Puente
Pe\~{n}alba, "The String Density of States from the Convolution
Theorem", {\em Phys. Lett.}{\bf B400} (1997) 275-283, hep-th/9701122;
"Counting Closed String States in a Box",  {\em Phys. Lett.}{\bf B408}
(1997) 183-191, hep-th/9705185.

\bibitem{blackholes1}
 T. Banks, W. Fischler, I. R. Klebanov and  L. Susskind,
"Schwarzschild Black Holes from Matrix Theory", {\em Phys. Rev. Lett. }{\bf
80} (1998) 226-229, hep-th/9709091. \\
I. R. Klebanov and  L. Susskind, "Schwarzschild Black Holes in Various
Dimensions from Matrix Theory", {\em  Phys. Lett.}{\bf B416 } (1998) 62-66,
hep-th/9709108. \\

\bibitem{blackholes2}
 T. Banks, W. Fischler, I. R. Klebanov and  L. Susskind,
"Schwarzschild Black Holes from Matrix Theory II" hep-th/9711005. \\    
 H. Liu and  A. A. Tseytlin, "Statistical Mechanics of D0-branes and Black
Hole Thermodynamics",   JHEP {\bf 01 }(1998) 010,  hep-th/9712063. \\
N.Ohta and J.Zhou "Euclidean Path Integral, D0-branes and Schwarzschild
Black Holes in Matrix Theory", hep-th/9801023\\
D.A.Lowe "Statistical origin of Black Hole Entropy", hep-th/9802173.

\bibitem{bb} K. Becker and M. Becker, "A Two-Loop Test of M(atrix) Theory",
 hep-th/9705091.

\bibitem{enrial} E. \'Alvarez, "Strings at Finite Temperature", {\it
Nucl. Phys.} {\bf B269} (1986) 596.

\bibitem{emar87} E. \'{A}lvarez and M. A. R. Osorio, "Superstrings at
Finite Temperature", {\em Phys. Rev} {\bf D36} (1987) 1175-1183.

\bibitem{danfer} M. Claudson and M. B. Halpern,
"Supersymmetric Ground State Wave Functions", {\it
Nucl. Phys.} {\bf B250} (1985) 689.  \\
 U. H. Danielsson, G. Ferretti and  B. Sundborg,
"D-particle Dynamics and Bound States", {\em  Int. J. Mod. Phys.}{\bf
A11}(1996)
5463-5478, hep-th/9603081. \\
 D. Kabat and  P. Pouliot,
"A Comment on Zero-brane Quantum Mechanics", {\em Phys. Rev. Lett.}{\bf
77} (1996) 1004-1007, hep-th/9603127.

\bibitem{kabtaylor} D.Kabat and W.Taylor "Linearized supergravity from Matrix
theory", hep-th/9712185.
\\ 
M. Raamsdonk  "Conservation of Supergravity Currents from Matrix
Theory",hep-th/9803003.
     
\bibitem{rutgers} M. R. Douglas, D. Kabat, P. Pouliot, S. H. Shenker,
"D-branes and Short Distances in String Theory", {\em Nucl. Phys. } {\bf
B485} (1997) 85-127, hep-th/9608024.

\bibitem{malda}C.G. Callan and J.Maldacena "D-Brane approach to black
hole quantum mechanics", {\em Nucl. Phys.}{\bf B 472} (1996)591,
hep-th/9602043

\bibitem{malstr}J.Maldacena and A.Strominger "Black hole greybody factors and
D-brane spectroscopy", {\em Phys. Rev.}{\bf D 55} (1997)861,
hep-th/9609026

\bibitem{tse}A. Tseytlin "Open superstring partition function in constant
gauge field background at finite temperature", hep-th/9802133

\bibitem{greenmavaz} M. A. V\'azquez-Mozo, "Open String Thermodynamics and
D-Branes"  {\em Phys. Lett.}{\bf B388 }(1996) 494, hep-th/9607052. \\
M. B. Green, {\em Nucl. Phys.  }{\bf B381} (1992)201 .

\bibitem{pol}J. Polchinski, "Dirichlet-Branes and Ramond-Ramond Charges", 
{\em Phys. Rev. Lett.}{\bf 75} (1995) 4724-4727, hep-th/9510017; "TASI
Lectures on D-Branes",hep-th/9611050.

\bibitem{kle} I. R. Klebanov and  A. A. Tseytlin,
"Entropy of Near-Extremal Black p-branes",{\em  Nucl. Phys.}{\bf B475 }
(1996) 164-178, hep-th/9604089. \\
S. S. Gubser, I. R. Klebanov and  A. W. Peet, "Entropy and Temperature
of
Black 3-Branes", {\em Phys. Rev.}{\bf D54 }(1996) 3915-3919,
hep-th/9602135.

\bibitem{bound}  E. Witten, "Bound States of Strings and p-Branes",
{\em Nucl. Phys.} {\bf  B460 }(1996) 335-350, hep-th/9510135.

\bibitem{moore} G. Moore, "Modular Forms and Two-Loop String Physics",
{\it Phys. Lett.} {\bf B176} (1986) 369-379.

\bibitem{taylor}
W. Taylor, "D-brane field theory on compact spaces",
{\em Phys. Lett.}{\bf B394} (1997) 283-287, hep-th/9611042;
"Lectures on D-branes, Gauge Theory and M(atrices)", PUPT-1762
and hep-th/9801182. 

\end{thebibliography}
\end{document}